\documentclass[preprint,showpacs,preprintnumbers,amsmath,amssymb]{revtex4}

\usepackage{graphicx}
\usepackage{dcolumn}
\usepackage{bm}

\begin{document}
\draft
\title{Superheavies: Theoretical incitements and predictions}

\author{V.I.~Zagrebaev, A.V. Karpov}
\affiliation{Flerov Laboratory of Nuclear Reactions, JINR, Dubna,
Moscow Region, Russia}
\author{I.N. Mishustin and Walter~Greiner}
\affiliation{Frankfurt Institute for Advanced Studies, Frankfurt am Main, Germany}

\begin{abstract}
It is well known that in fusion reactions one may get only neutron deficient
superheavy nuclei located far from the island of stability.
The multi-nucleon transfer reactions allow one to produce more neutron enriched
new heavy nuclei but the corresponding cross sections are rather low.
Neutron capture process is considered here as alternative method for production
of long-lived neutron rich superheavy nuclei. Strong neutron fluxes might be provided by nuclear reactors
and nuclear explosions in laboratory frame and by supernova explosions in nature.
All these cases are discussed in the paper.
\end{abstract}

 \maketitle

\section{Motivation}\label{In}

A ten years epoch of $^{48}$Ca irradiation of actinide targets for the synthesis of superheavy (SH) elements is over. The heaviest available targets
of Berkelium ($Z=97$) and Californium ($Z=98$) have been used to produce the elements 117 \cite{Ogan10} and 118 \cite{Ogan06}.
Note that the predicted cross sections and excitation functions \cite{Z03,Z04} have been fully confirmed
by the experiments performed in Dubna and later in Berkeley and GSI. These predictions are based on the shell corrections
to the ground states of superheavy nuclei calculated by P.~M\"oller and others \cite{Moller95} which we used to estimate survival probabilities
of these nuclei against fission. It is the gradual increase of the fission barriers of the compound nuclei (CN) formed in these reactions which
explains the almost constant value (of a few picobarns) of the cross sections for the production of SH elements with $Z=112\div 118$
in hot fusion reactions.

\begin{figure}[ht]
\includegraphics[width=1.0\textwidth]{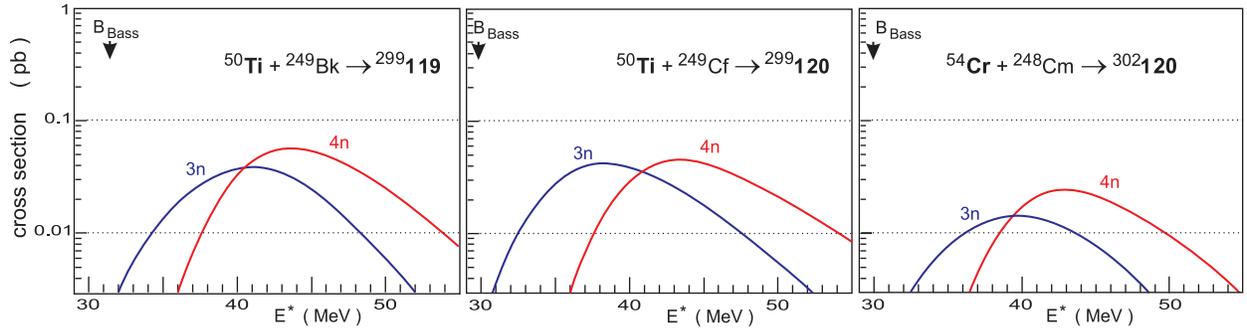}
\caption{Excitation functions for production of superheavy elements 119 and 120
in the 3n and 4n evaporation channels of the $^{50}$Ti+$^{249}$Bk, $^{50}$Ti+$^{249}$Cf
and $^{54}$Cr+$^{248}$Cm fusion reactions.\label{120}}
\end{figure}

As mentioned above, $^{249}$Cf (T$_{1/2}$ = 351 yr) is the heaviest available target that can be used in experiment.
Thus, to get SH elements with $Z > 118$ in fusion reactions, one should proceed to heavier than $^{48}$Ca projectiles.
The strong dependence of the calculated evaporation residue (EvR) cross sections for the production of element 120 on the mass asymmetry
in the entrance channel \cite{Zag08} makes the nearest to $^{48}$Ca projectile, $^{50}$Ti, most promising for further
synthesis of SH nuclei. Of course, the use of the titanium beam instead of $^{48}$Ca also decreases the yield of SH nuclei
mainly due to a worse fusion probability. The calculated excitation functions for the synthesis of SH elements 119 and 120
in the fusion reactions of $^{50}$Ti with $^{249}$Bk and $^{249}$Cf targets are shown in Fig.\ \ref{120} taken from Ref.\cite{Zag08}.

The orientation effects are known to play an important role in fusion reactions of statically deformed heavy nuclei \cite{Z03,Z04}.
The fusion probability (formation of CN) is strongly suppressed for more elongated nose-to-nose initial orientations.
As a result, the preferable beam energies for the synthesis of SH elements in the hot fusion reactions are shifted to values that
are several MeV higher than the corresponding Bass barriers (calculated for spherical nuclei).
As can be seen from Fig.\ \ref{120}, the estimated EvR cross sections for the 119 and 120 SH elements synthesized in the $^{50}$Ti
induced fusion reactions are quite reachable at available experimental setups, though one needs much longer time of irradiation than for
the $^{48}$Ca fusion reactions.

The yield of superheavy nuclei (number of events per day) depends not only on the cross section but also
on the beam intensity and target thickness. In this connection the other projectile--target combinations should be also considered.
Most neutron-rich isotopes of element 120 may be synthesized in the three different fusion reactions $^{54}$Cr+$^{248}$Cm,
$^{58}$Fe+$^{244}$Pu and $^{64}$Ni+$^{238}$U leading to the same SH nucleus $^{302}120$ with neutron number near to the predicted closed
shell $N=184$ \cite{MorelGreiner1969}. These three combinations are not of equal value. The estimated EvR cross sections for more symmetric $^{58}$Fe+$^{244}$Pu
and $^{64}$Ni+$^{238}$U reactions are lower than those of the less symmetric $^{54}$Cr+$^{248}$Cm combination \cite{Zag08}, which,
in its turn, is quite comparable with the Ti-induced fusion reaction, see the right panel of Fig.\ \ref{120}.
The advantage factor 2 or 3 for the $^{50}$Ti+$^{249}$Bk and $^{50}$Ti+$^{249}$Cf fusion reaction as compared with $^{54}$Cr+$^{248}$Cm
is definitely within the theoretical uncertainty for calculation of such small cross sections.

Such uncertainty originates from two main factors. First, it is connected with the complex dynamics of CN formation in competition
with the dominating quasi-fission process. There are too few experimental data on low-energy collisions of very heavy ions to fix finally
all the parameters (potential energy, friction forces, nucleon transfer rate) and to obtain an explicit value for the probability of CN
formation for such heavy systems. The second source of uncertainty comes from the estimation of the survival probability of the excited CN,
which is calculated within the statistical model. The basic quantities here are the fission barriers of SH nuclei appearing
in the neutron evaporation chain. For example, in the 4n evaporation channel one needs to know the fission barriers and neutron separation energies
for 5 subsequent isotopes of SH element.

\begin{figure}[ht]
\includegraphics[width=1.0\textwidth]{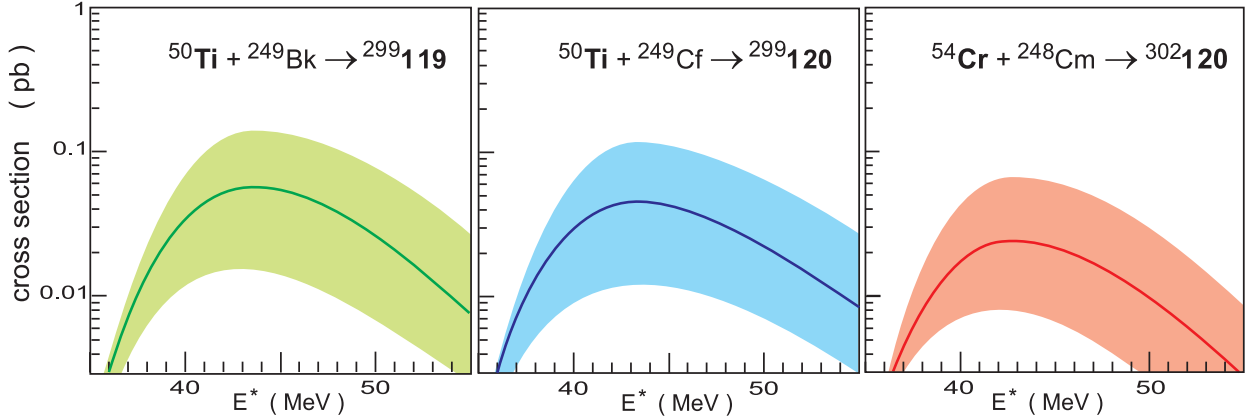}
\caption{Excitation functions for the production of superheavy elements 119 and 120 in the 4n evaporation channels of the $^{50}$Ti+$^{249}$Bk,
$^{50}$Ti+$^{249}$Cf and $^{54}$Cr+$^{248}$Cm fusion reactions. A factor 3 is assumed for the uncertainty of the survival probability
caused by the uncertainty of the fission barriers of SH nuclei.\label{TiCr}}
\end{figure}

Different theoretical models give quite different values for the fission barrier for heavy nuclei.
As mentioned above, we used the fission barriers of SH nuclei predicted by the macro--microscopic model \cite{Moller95},
which gives lower fission barriers for the isotopes of elements 119 and 120 than for elements $114\div 116$ synthesized in $^{48}$Ca
fusion reactions. On the other hand, the full microscopic models based on self-consistent Hartree-Fock calculations usually predict
much higher fission barriers for these isotopes (up to 10 MeV) if the Skyrme forces are used \cite{Burv04} (though these predictions
are not unambiguous and depend strongly on chosen nucleon-nucleon forces). The exponential dependence of the survival probability
on the values of the fission barriers leads to a change of it approximately by a factor 10 if the barriers are changed by 1 MeV.

Thus, we may conclude that the theoretical uncertainty in the predictions of the cross sections for formation of SH nuclei in hot fusion
reactions with actinide targets is not less than a factor 3. If we consider such an uncertainty in our calculations, the excitation functions
for synthesis of the new SH elements 119 and 120 in 4n evaporation channels of the $^{50}$Ti+$^{249}$Bk, $^{50}$Ti+$^{249}$Cf and $^{54}$Cr+$^{248}$Cm
fusion reactions will look as shown in Fig.\ \ref{TiCr}. From this figure it is clear that all these reactions must be considered as quite
promising for the synthesis of new elements. The final choice between them depends not so much on the difference in the cross sections
as on other experimental conditions (availability of appropriate targets, beam intensities, etc.).

\begin{figure}[ht]
\begin{center} \resizebox*{9.5 cm}{!} {
\includegraphics{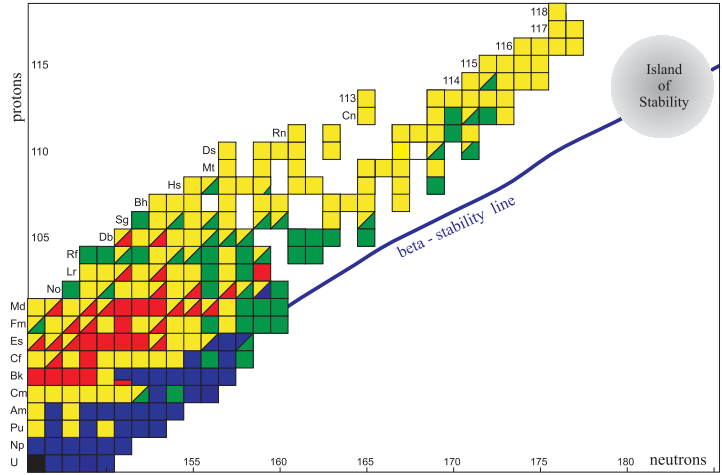}} \end{center}
\caption{Upper part of the nuclear map. The island of stability is shown schematically.\label{map_up}}
\end{figure}

Note, however, that the present limits for neutron rich isotopes located at the upper part of the nuclear
map ($Z > 60$) are very close to the line of stability while the unexplored area
of heavy neutron-rich nuclides (to the east of the stability line) is extremely
important for nuclear astrophysics investigations and, in particular,
for the understanding of the r-process of astrophysical nucleogenesis.
For nuclei with $Z > 100$ only neutron deficient isotopes (located to the left of the
stability line) have been synthesized so far, see Fig.\ \ref{map_up}. Due to the bending of the stability line to the neutron axis,
in fusion reactions of stable nuclei one may produce only proton rich isotopes of heavy elements
(it's true also for the isotopes of 119 and 120 elements discussed above).
This is the main reason for the impossibility of reaching the center of the ``island of stability''
in the superheavy mass region ($Z\sim 114\div 120$ and $N \sim 184$) in fusion reactions with stable
projectiles. In the $^{50}$Ti+$^{249}$Cf and $^{54}$Cr+$^{248}$Cm fusion reactions the 120-th element
can be synthesized with cross sections of about 0.04 and 0.02 pb, correspondingly \cite{Zag08},
but again only the short-lived neutron deficient isotopes of this element may be obtained.
The use of beams of radioactive nuclei hardly may solve this problem due
to their low intensities.

Multi-nucleon transfer processes in low-energy collisions of actinide nuclei (like U+Cm)
may really lead to the formation of neutron rich long-living superheavies. The shell effects
(antisymmetrizing quasi-fission process with a preferable formation of nuclei close to
the doubly magic lead isotope) might significantly enhance the corresponding
cross sections \cite{Zag06,Zag08}. In spite of difficulties of separation of the transfer
reaction products, experiments of such kind are planned to be performed in nearest future.

\begin{figure}[ht]
\begin{center} \resizebox*{8.0 cm}{!} {
\includegraphics{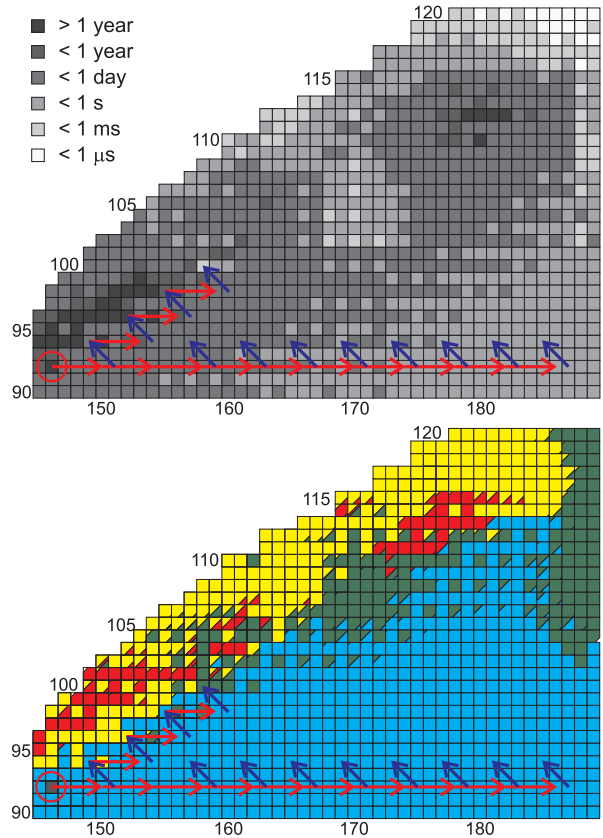}} \end{center}
\caption{Half-lives (up) and decay modes (bottom) of nuclei in the upper part of the nuclear map.
Schematic view of slow (terminated at the short-lived fission Fermium isotopes) and fast neutron capture processes with
subsequent beta-minus decays are shown by the arrows.\label{maps}}
\end{figure}

The neutron capture process is an alternative (oldest and natural) method for the production
of new heavy elements. Strong neutron fluxes might be provided by nuclear reactors
and nuclear explosions under laboratory conditions and by supernova explosions in nature.
However the ``Fermium gap'', consisting of the short lived Fermium isotopes $^{258-260}$Fm
located at the beta stability line while have very short half-lives for spontaneous fission,
impedes formation of nuclei with Z$>$100 by the weak neutron fluxes realized in existing
nuclear reactors. In nuclear and supernova explosions (fast neutron capture) this gap may
be bypassed, if the total neutron fluence is high enough.
Theoretical models predict also another region of short lived nuclei located at Z=106$\div$108
and A$\sim$270 (see Fig.\ \ref{maps}).

In this paper we study the possibility of synthesizing heavy elements in multiple ``soft'' nuclear
explosions and in pulsed reactors. We have found that in the first case the both gaps may be easily bypassed and,
thus, a measurable amount of the neutron rich long-lived superheavy nuclei of the island of stability
may be synthesized. For the second case we have formulated requirements for the pulsed reactors of the next generation
which could be also used for the production of long-lived superheavy nuclei.

\section{Nucleosynthesis by neutron capture}\label{nflux}

The synthesis of heavier nuclei in the reaction of neutron capture with subsequent beta-minus decay
is a well studied process (see, e.g., \cite{Dorn62,Seaborg,Kriv69}).
Relative yields of the isotopes formed in such a process may be found as a solution
of the following set of differential equations (somewhat simplified here)

\begin{eqnarray}\label{equation}
&\displaystyle {\frac{dN_{Z,A}}{dt}} = N_{Z,A-1} n_0 \sigma_{n\gamma}^{Z,A-1} - N_{Z,A} n_0 \sigma_{n\gamma}^{Z,A}
\\ \nonumber & - N_{Z,A}[\lambda_{Z,A}^{\beta -} + \lambda_{Z,A}^{fis} + \lambda_{Z,A}^{\alpha}]
+ N_{Z-1,A}\lambda_{Z-1,A}^{\beta -} + N_{Z+2,A+4}\lambda_{Z+2,A+4}^{\alpha},
\end{eqnarray}
where $n_0$ is the neutron flux (number of neutrons per square centimeter per second) and
$\lambda_{Z,A}^i=ln2/T^i_{1/2}$ is the decay rate of the nucleus $(Z,A)$ into the channel $i$
(i.e., beta-minus and alpha decays and fission).
For simplicity we ignore here the energy distribution of the neutrons and, thus, the energy dependence
of the neutron capture cross section $\sigma_{n\gamma}^{Z,A}$. Neutrons generated by fission
in nuclear reactors and in explosions are rather fast (far from the resonance region).
In the interval of 0.1--1~MeV the neutron capture cross section is a smooth function of energy with the value
of about 1 barn, which is used below for numerical estimations. Note that integration over the neutron energy
may be performed very easily and does not change the obtained conclusions. The simplest version
of the reaction chain, where only the neutron capture reactions are retained [first line in Eq.(\ref{equation})],
was considered recently in Ref.\cite{Botvina10}.

To solve Eq.(\ref{equation}) numerically one needs to know the decay properties of neutron rich
nuclei which are not studied yet. That is the main problem, which significantly complicates
the analysis of the multiple neutron capture processes.
Theoretical estimations of half-lives for $\alpha$-decay are rather reliable because they
depend only on ground state masses which are very close in different theoretical models.
For the half-lives of $\alpha$-decays we used the well-known Viola-Seaborg formula
with the coefficient proposed by Sobiczewski et al. \cite{Sobiczewski1989}
Half-lives of allowed $\beta$-decays depend also on the ground state masses of the nuclei and may be
estimated by the empirical formula $\log_{10}{\left[f_0 T_{1/2}^{\beta}\left({\rm sec}\right)\right]}= 5.7$,
where the Fermi function $f_0$ is calculated by the standard formulas (see, e.g., Ref.~\cite{Wu}).

The fission half-lives are the most uncertain quantities in Eqs.(\ref{equation}).
Reliable analysis of the fission process requires knowledge of the multidimensional potential energy
surface as well as of the collective inertia parameters. Such calculations can be performed
only in very restricted areas of the nuclear map.
We used the ground state masses and shell corrections for heavy and superheavy nuclei
proposed by P.~M\"{o}ller et al.\cite{Moller95} and then we applied the empirical formula for the estimation
of the fission half-lives. For this purpose we employed the relation by Swiatecki \cite{Swiatecki1955}
based on the idea of the dominant role of the fission barrier for the fission probability
\begin{eqnarray}\label{T_SF}
&&\log_{10}{T_{1/2}^{fis}\left({\rm sec}\right)} = 1146.44 - 75.3153
Z^2/A
+1.63792 \left(Z^2/A\right)^2  \nonumber\\&&- 0.0119827 \left(Z^2/A\right)^3
+B_f \left(7.23613 - 0.0947022 Z^2/A\right)+\nonumber\\
&&+\left\{
\begin{array}{cl}
  0,       & \mbox{Z and N are even} \\
  1.53897, & \mbox{A is odd} \\
  0.80822, & \mbox{Z and N are odd} \\
\end{array}
\right.
\end{eqnarray}
Here $B_f=B_f^{LDM} + \delta U_{g.s.}$ is the fission barrier, which was calculated as a sum of the
liquid-drop barrier and the ground-state shell correction.
The coefficients of the systematics (\ref{T_SF}) were determined by the fitting to the experimental data
and to the rather realistic theoretical predictions \cite{Smolanczuk1997,Pomorski1996} for the region
of $100\le Z\le 120$ and $140\le N\le 190$.

The results of such calculations
are shown in the left part of Fig.\ \ref{maps}, where the theoretical values of
$T_{1/2}^{\alpha}$, $T_{1/2}^{\beta}$ and/or $T_{1/2}^{fis}$ were replaced by experimental ones
if known. In accordance with our calculations, the most stable superheavy nuclei located
at the island of stability (which could be searched for in nature) are the isotopes of
Darmstadtium ($Z=110$) and/or Copernicium ($Z=112$) with the neutron number $N\sim 180$.
To produce these nuclei in the neutron capture process one needs to bypass the two area of short-living
fissile nuclei, namely, the Fermium gap ($Z=100$) and the region of Z=106$\div$108 and A$\sim$270.

To test our model we first described available data on the fast neutron capture process
realized in nuclear explosions. In this case the time of neutron capture,
$\tau_n = (n_0\sigma_{n\gamma})^{-1}\sim 1$~$\mu$s $<<T_{1/2}(Z,A)$, is much shorter
than half-lives of the produced nuclei (up to the neutron drip line).
Keeping only the first two terms of the r.h.s. of Eq.(\ref{equation}),
we get the following analytical solution (with initial conditions
 $N_{Z,A}(t=0)=1$ and $N_{Z,A+k}(t=0)=0$ at $k>0$ where $k$ is the number of captured
 neutrons)
\begin{equation}\label{sol}
N_{Z,A+k}=\frac{x^k}{k!}e^{-x}.
\end{equation}
This relation can be used for clear understanding of the fast neutron capture process and for
a preliminary estimation of relative yields of heavy nuclei synthesized in such a process.
Here $x=n\sigma_{n\gamma}$, $n=n_0\tau$ is the total neutron fluence (neutrons per
square centimeter) and $\tau$ is the duration of explosive neutron irradiation. Thus, the dimensionless quantity
$x=n\sigma_{n\gamma}$ is a key factor characterizing the neutron capture process. In nuclear explosions
the neutron fluence reaches the values of $10^{25}$~cm$^{-2}$, so that $x\sim 1$ and more than 10 neutrons
could be captured during one exposure with time duration of about $1\,\mu$s \cite{Seaborg}.

\begin{figure}[ht]
\begin{center} \resizebox*{8.0 cm}{!} {
\includegraphics{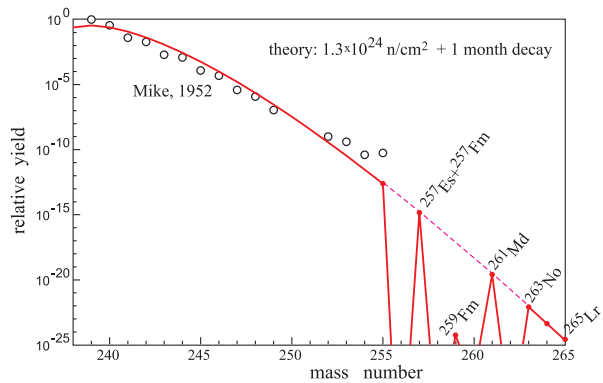}} \end{center}
\caption{Experimental (open dotes) and calculated relative yields of heavy nuclei
in the test nuclear explosion ``Mike''\cite{Mike}. \label{mike}}
\end{figure}

In Fig.\ \ref{mike} the experimental data on the yield of transuranium nuclei in the test
thermonuclear explosion ``Mike'' \cite{Mike} are compared with those calculated by Eqs.(\ref{equation})
assuming $1\,\mu$s neutron exposure of $1.3\times 10^{24}$~neutrons/cm$^2$ with subsequent
one-month decay time. Note that elements 99 and 100 (Einsteinium and Fermium) were first discovered
just in debris of the ``Mike'' explosion. As can be seen, in this case the Fermium gap does not
influence the yields of nuclei with $Z>100$ which follow roughly the relation (\ref{sol}).

\begin{figure}[h]\begin{center}
\resizebox*{7.0 cm}{!}
{\includegraphics{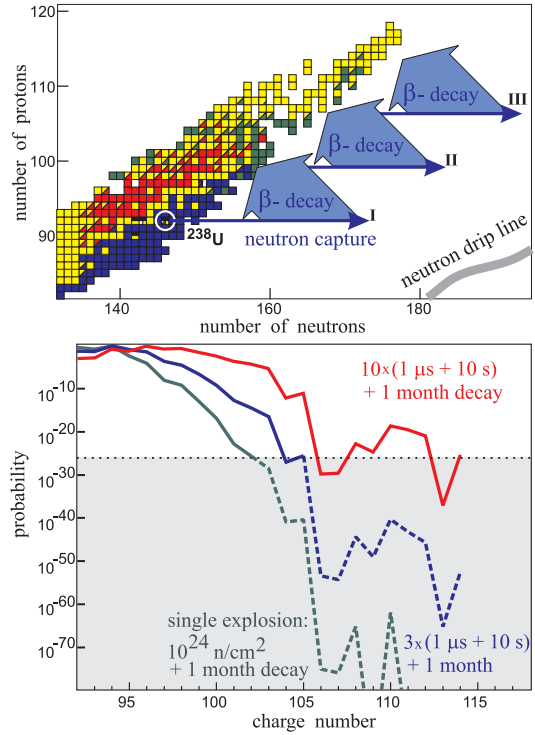}}\end{center}
\caption{Schematic picture for multiple neutron irradiation of initial $^{238}$U material (up)
and probability for formation of heavy nuclei (bottom) in such process (one, three and ten subsequent explosions).
Dotted line denotes the level of few atoms.\label{multiple}}
\end{figure}

Recently we proposed to consider the possibility of generating two or several nuclear
explosions in close proximity of each other to increase the resulting mass number
of the synthesized nuclei\cite{Botvina10}.
Here we study for the first time such a possibility illustrated in the upper part of Fig.\ \ref{multiple}.
In the bottom part of this figure the probabilities of heavy element formation are shown for one,
three and ten subsequent short-time ($1\,\mu$s) neutron exposures of $10^{24}$~n/cm$^2$ each
following one after another with time interval of 10 seconds with final one month waiting
(needed to perform some experimental measurements).

Our results demonstrate for the first time that multiple rather ``soft'' nuclear explosions
could be really used for the production of noticeable (macroscopic) amount of neutron rich long-lived
superheavy nuclei. Leaving aside any discussions on possibilities of such processes
and associated technical problems, we want to emphasize a sharp increase of the probability
for formation of heavy elements with $Z\ge 110$ in the multiple neutron irradiations
(enhancement by several tens of orders of magnitude, see Fig.\ \ref{multiple}).
This probability is high enough for some superheavy elements (see the region above the dotted line
in Fig.\ \ref{multiple}) to perform their experimental identification.

\begin{figure}[h]\begin{center}
\resizebox*{8.0 cm}{!}
{\includegraphics{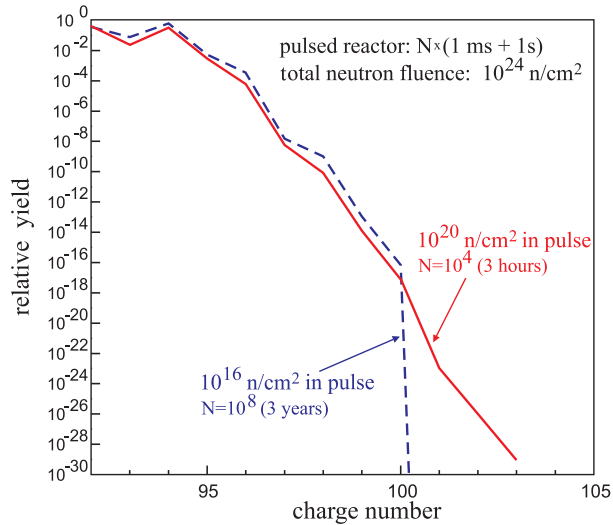}}\end{center}
\caption{Relative yields of heavy elements in ordinary (dashed line) and in high--intensity
pulsed reactors (solid line) at the same total neutron fluence $10^{24}$ n/cm$^2$.\label{reactor}}
\end{figure}

It would be interesting also to study the same process of multiple neutron exposures
realized in pulsed nuclear reactors. The pulse duration here could be much longer
than in nuclear explosions (up to few milliseconds). In spite of that, the neutron fluence
usually does not exceed $10^{16}$ n/cm$^2$ in existing nuclear reactors ($n_0\sim 10^{19}$~n/cm$^2$s during
one millisecond pulse). Thus, the quantity $x=n\sigma_{n\gamma}$ is about $10^{-8}$,
the time of neutron capture $\tau_n = (n_0\sigma_{n\gamma})^{-1}\sim 10^5$~s,
and only the nearest long lived isotopes (A+1 or A+2) of irradiated elements may be formed during the pulse, see formula (\ref{sol}).
Multi--pulse irradiation here corresponds, in fact, to the ``slow'' neutron capture process,
in which new elements with larger charge numbers are situated close to the line of stability
and finally reach the Fermium gap where the process stops (see Fig.\ \ref{maps}). The result of numerical solution
of Eq.(\ref{equation}) for neutron capture process in an ordinary pulsed reactor is shown
in Fig.\ \ref{reactor} by the dashed line. The probability for formation of heavy elements
with $Z>100$ is negligibly small here independent of a number of pulses and total time
of irradiation.

The situation may change if one would be able to increase somehow the intensity of the pulsed reactor.
The neutron fluence in one pulse and frequency of pulses should be high enough to bypass the both gaps
of short-lived nuclei on the way to the island of stability (see left part of Fig.\ \ref{maps}).
Thus, the specification of the high--intensity pulsed reactors of next generation depends strongly
on properties of heavy neutron rich nuclei located to the right of these gaps. These nuclei are not
discovered yet (see Fig.\ \ref{map_up}), and undoubtedly certain experimental efforts should be made
to resolve this problem.

Using our theoretical estimations for the decay properties of these nuclei (see above) we have found that
increase of the neutron fluence in the individual pulse by about three orders of magnitude as compared with existing
pulsed reactors, i.e. up to 10$^{20}$ netrons/cm$^2$, could be quite sufficient to bypass the both gaps
(see the solid curve in Fig.\ \ref{reactor}). The details of the calculations and estimations of
the yields of long-lived superheavy nuclei produced in neutron capture processes will be published elsewhere.

\section{Summary}\label{sm}

We have shown for the first time that a macroscopic amount of the long-living superheavy nuclei located
at the island of stability may be really produced in multiple (rather ``soft'') nuclear explosions, if such
processes would be realized technically. This goal could be also reached by using the pulsed nuclear reactors
of next generation, if their neutron fluence per pulse will be increased by about three orders of magnitude.
The experimental study of the decay properties of heavy nuclei located along the beta-stability line
(and to the right of it, see Fig.\ \ref{map_up}) is extremely important for a more accurate analysis of the
neutron capture processes (including the astrophysical ones \cite{Panov09}) in the upper part of the nuclear map.

We are indebted to the DFG -- RFBR collaboration and to GSI for support of our studies.


\begin{thebibliography}{99}

\bibitem{Ogan10} Yu.Ts.~Oganessian et al., {\it Phys. Rev. Lett.} {\bf 104}, 142502 (2010).

\bibitem{Ogan06} Yu.Ts.~Oganessian, V.K.~Utyonkov et al., {\it Phys. Rev.} C{\bf 74}, 044602 (2006).

\bibitem{Z03} V.I.~Zagrebaev, M.G.~Itkis, and Yu.Ts.~Oganessian, {\it Phys. At. Nucl.} {\bf 66}, 1033 (2003).

\bibitem{Z04} V.I.~Zagrebaev, {\it Nucl. Phys.} A{\bf 734}, 164 (2004).

\bibitem{Moller95} P.~M\"oller, J.~R.~Nix, W.~D.~Myers, W.~J.~Swiatecki,
{\it At. Data Nucl. Data Tables} {\bf 59}, 185 (1995).

\bibitem{Zag08} V.I.~Zagrebaev and W.~Greiner, {\it Phys. Rev.} C {\bf 78}, 034610 (2008).

\bibitem{MorelGreiner1969} U. Mosel and W. Greiner, {\it Z. Phys.} {\bf 222}, 261 (1969); see also {\it Memorandum zur Errichtung eines gemeinsamen
Ausbildungszentrums f\"ur Kernphysik Hessischer Hochschulen} (1966).

\bibitem{Burv04} T.~B\"{u}rvenich, M.~Bender, J.A.~Maruhn, and P.-G.~Reinhard, {\it Phys. Rev.} C {\bf 69}, 014307 (2004).

\bibitem{Zag06} V.I.~Zagrebaev, Yu.Ts.~Oganessian, M.G.~Itkis and W.~Greiner,
{\it Phys. Rev.} C {\bf 73}, 031602 (2006).

\bibitem{Dorn62} D.W.~Dorn, {\it Phys. Rev.} {\bf 126}, 693 (1962).

\bibitem{Seaborg} G.T.~Seaborg, {\it Ann. Rev. Nucl. Sci.}, {\bf 18}, 119 (1968).

\bibitem{Kriv69} A.S.~Krivochatsky, Yu.F.~Romanov, {\it Production of transuranium
and actinide elements by neutron irradiation (in Russian)}, Atomizdat, Moscow, 1969.

\bibitem{Botvina10} A.~Botvina, I.~Mishustin, V.~Zagrebaev and
Walter Greiner, {\it Int. J. Mod. Phys.} E{\bf 19}, 2063 (2010).

\bibitem{Sobiczewski1989} A.~Sobiczewski, Z.~Patyk, and S.~\v{C}woik,
{\it Phys. Lett.} {\bf B224}, 1 (1989).

\bibitem{Wu} C.S.~Wu and S.A.~Moszkowski,
{\it Beta decay} (John Wiley \& Sons, New York, 1966), p. 183.

\bibitem{Swiatecki1955} W.J.~Swiatecki, {\it Phys. Rev.} {\bf 100}, 937 (1955).

\bibitem{Smolanczuk1997} R.~Smola\'{n}czuk, {\it Phys. Rev.} {\bf C56}, 812 (1997).

\bibitem{Pomorski1996} A.~Staszczak,  Z.~{\L}ojewski, A.~Baran, B.~Nerlo-Pomorska, K.~Pomorski, in {\it Dynamical Aspects
of Nuclear Fission}, Casta-Papernicka, (1996), 22.

\bibitem{Mike} H.~Diamond et al., {\it Phys. Rev.} {\bf 119}, 2000 (1960).

\bibitem{Panov09} I.V.~Panov, I.Yu.~Korneev and F.-K.~Thielemann, {\it Phys. At. Nucl.} {\bf 72}, 1026 (2009).

\end{thebibliography}
\end{document}